\documentclass[aps,prd,preprint,floats,epsf,superscriptaddress]{revtex4}
\usepackage{graphicx}
\usepackage{amssymb}
\usepackage{amsfonts}
\usepackage{amsmath}

\def\bwt{\begin{widetext}}
\def\ewt{\end{widetext}}
\def\be{\begin{equation}}
\def\ee{\end{equation}}
\def\bea{\begin{eqnarray}}
\def\eea{\end{eqnarray}}
\def\bean{\begin{eqnarray*}}
\def\eean{\end{eqnarray*}}
\def\bary{\begin{array}}
\def\eary{\end{array}}

\def\bit{\begin{itemize}}
\def\eit{\end{itemize}}

\begin{document}

%\preprint{\vbox{ \hbox{hep-ph/yymmnnn}
%         }}
\bigskip

\title{A six-dimensional gauge-Higgs unification model based on $E_6$ gauge
  symmetry}

\author{Cheng-Wei Chiang}
\email[e-mail: ]{chengwei@ncu.edu.tw}
\affiliation{Center for Mathematics and Theoretical Physics and Department of
  Physics, National Central University, Chungli, Taiwan 320, R.O.C.}
\affiliation{Institute of Physics, Academia Sinica, Taipei, Taiwan 115, R.O.C.}
\author{Takaaki Nomura}
\email[e-mail: ]{nomura@ncu.edu.tw}
\affiliation{Center for Mathematics and Theoretical Physics and Department of
  Physics, National Central University, Chungli, Taiwan 320, R.O.C.}

\date{\today}

\begin{abstract}

  We construct a six-dimensional gauge-Higgs unification model with the enlarged gauge group of $E_6$ on $S^2/Z_2$ orbifold compactification.  The standard model particle contents and gauge symmetry are obtained by utilizing a monopole background field and imposing appropriate parity conditions on the orbifold.  In particular, a realistic Higgs potential suitable for breaking the electroweak gauge symmetry is obtained without introducing extra matter or assuming an additional symmetry relation between the SU(2) isometry transformation on the $S^2$ and the gauge symmetry.  The Higgs boson is a KK mode associated with the extra-dimensional components of gauge field.  Its odd KK-parity makes it a stable particle, and thus a potential dark matter candidate in the model.   We also compute the KK masses of all fields at tree level.

\end{abstract}


\maketitle

%%%%%%%%%%%%%%%%%%%%

\section{Introduction \label{sec:intro}}

Precision electroweak measurements suggest that the mass of the Higgs boson in the
standard model (SM) should be of ${\cal O}(100)$ GeV.  However, this leads to a
serious problem, the so-called hierarchy problem, because the Higgs boson mass
generally suffers from quadratic divergence at the quantum level.  It is thus
unnatural for the Higgs boson to be so light if the theory cutoff scale is high,
unless some mechanism is introduced for stabilization.  Such a problem generally calls for some symmetry ({\it e.g.}, supersymmetry) to control the scalar sector and leads to physics beyond the SM.

In the late 70's, an alternative method to stabilize the Higgs boson mass had been proposed.  The basic idea was to embed the Higgs field as the extra-dimensional components of gauge field in a higher dimensional space, with an enlarged gauge symmetry broken down to the SM gauge group in 4D spacetime \cite{Manton:1979kb,Forgacs:1979zs,Fairlie:1979at}.  This idea of gauge-Higgs unification had recently been revived \cite{Dvali:2001qr,ArkaniHamed:2001nc,Csaki:2002ur,Scrucca:2003ra,Antoniadis:2001cv,Hall:2001zb,%
  Burdman:2002se,Haba:2002vc,Choi:2003kq, Sakamura:2007qz, Medina:2007hz, Lim:2007jv, Hosotani:2008tx}.  A desirable feature of such models is that the gauge origin ensures that the Higgs mass in the bulk is protected from quadratic divergence.  Moreover, by compactifying the theory on orbifolds, unwanted fields can be projected out from low-energy spectrum. The compactification scale would be taken as TeV scale \cite{Antoniadis:1990ew, Antoniadis:1993jp}. A simple implementation of the idea in 5D, however, encounters the difficulty of a small Higgs mass due to the absense of a tree-level Higgs potential.  One is then led to consider 6D models because a quartic Higgs interaction term can arise from the gauge kinetic term \cite{Scrucca:2003ut}.  The Higgs mass can also be enhanced through the introduction of a warped spacetime \cite{Contino:2003ve, Hosotani:2005nz} or by choosing a suitable bulk matter content \cite{Haba:2004qf}.  However, the quadratic mass term here is still radiatively generated and possibly divergent.  A more successful 6D model based on the SO(12) gauge group was proposed, where a monopole background exists to break the higher dimensional symmetry and results in a negative squared mass \cite{Nomura:2008sx}.  Nevertheless, a set of symmetries relating the SU(2) isometry transformation on $S^2$ to the gauge transformation of the gauge fields has to be imposed in order to carry out dimensional reduction of the gauge sector.  This approach of dimensional reduction is known as the coset space dimensional reduction, and leads to a stronger constraint on the four dimensional Lagrangian after dimensional reduction \cite{Manton:1979kb, Forgacs:1979zs,Kapetanakis:1992hf}.

We consider a gauge-Higgs unification model defined on the 6D spacetime where the extra spatial dimensions are compactified on a 2-sphere $S^2$.  The gauge symmetry in the model constructed here is assumed to be $E_6$.  With a background field configuration and suitable boundary conditions of $S^2$ on the fields, we obtain the full SM particle contents as the zero modes in the model.  In particular, no relation between extra-dimensional isometry and gauge symmetries is needed.  We are able to identify a Higgs boson doublet coming from the two extra-spatial components of the gauge fields in the adjoint representation.  Unwanted modes are either projected out by compactification or given masses due to the interaction with the background field.  The Higgs potential in the effective 4D theory has the desired form to break the electroweak symmetry.  The compactification scale is fixed with the input of the $W$ boson mass.  A mass relation between the Higgs and $W$ bosons is obtained.  The Weinberg angle is the same as the usual SU(5) grand-unified theory (GUT).  Moreover, the Higgs particle is a Kaluza-Klein (KK) mode with an odd KK-parity.  It is stable under a $Z_2$ symmetry and thus a potential dark matter candidate.  Discussions about the dark matter candidate are also given in other gauge-Higgs unification models \cite{Panico:2008bx, Carena:2009yt, Hosotani:2009jk, Haba:2009xu}.

This paper is organized as follows.  In Section~\ref{sec:model}, we describe the 6D model compactified on the $S^2/Z_2$ orbifold.  A consistent set of parity assignments of fields in both representations is given, followed by reviewing the branching of the $E_6$ group and the reduction of its fundamental and adjoint representations.  We then work out the details of obtaining the SM particle contents as the zero modes of gauge and fermion fields in the model.  In Section~\ref{sec:higgs}, we identify a KK mode of an appropriate representation of the extra-dimensional components of gauge field as the Higgs field in the SM.  After obtaining the required commutation relations of gauge generators, we compute the tree-level Higgs potential.  The result is then used to obtain a relation between the Higgs mass and the $W$ boson mass.  In Section~\ref{KKmass}, we discuss the KK mode mass spectra for fermions and gauge bosons in the existence of the background gauge field.  We find that the Higgs boson in the model is a potential candidate of dark matter due to its odd KK-parity.  Our findings are summarized in Section~\ref{sec:summary}.

\section{Model \label{sec:model}}

In this section, we develop the model based on E$_6$ gauge symmetry in six-dimensional spacetime with $S^2/Z_2$ extra space.  On the orbifold $S^2/Z_2$, a set of non-trivial boundary conditions is imposed to restrict the gauge symmetry and massless particle contents in four-dimensional spacetime.  We also introduce in this model a background gauge field, which corresponds to a Dirac monopole configuration, to obtain chiral fermions in four dimensions.  We then show how the E$_6$ gauge symmetry is reduced to the SM gauge symmetry with some extra U(1)'s, {\it i.e.}, SU(3) $\times$ SU(2) $\times$ U(1)$_Y$ $\times$ U(1)$_X$ $\times$ U(1)$_Z$, and how the massless gauge bosons and the SM Higgs boson in four dimensions are obtained in the model.  We note in passing that all gauge groups of lower ranks ({\it e.g.}, SO(10), SO(11), SU(6), etc) either cannot give a Higgs field in the right representation or do not support SM chiral fermions in four dimensions.

%%%%%%%%%%%%%%%%%%%%%%%%%%%%%%%%%%%%%%%%%%%%%%%%%%%%%%%%%%%%%%%%%%%%%%%%%%%%%%%%%%%%%%%%

\subsection{Action in six-dimensional spacetime}

We start by considering the E$_6$ gauge symmetry group in six-dimensional spacetime, which is assumed to be a direct product of the four-dimensional Minkowski spacetime $M^4$ and the compactified two-sphere orbifold $S^2/Z_2$, {\it i.e.}, $M^4 \times S^2/Z_2$.  The two-sphere has a radius of $R$.  We denote the six-dimensional spacetime coordinates by $X^M = (x^{\mu}, y^{\theta}=\theta, y^{\phi}=\phi)$, where $x^{\mu}$ and $\{ \theta, \phi \}$ are the $M^4$ coordinates and spherical coordinates of $S^2$, respectively.  The spacetime index $M$ runs over $\mu \in \{0,1,2,3 \}$ and $\alpha \in \{ \theta, \phi \}$.  The orbifold $S^2/Z_2$ is defined by the identification of $(\theta,\phi)$ and $(\pi - \theta,-\phi)$ \cite{Maru:2009wu}.  The two fixed points are $(\pi/2,0)$ and $(\pi/2,\pi)$.  The spacetime metric of $M^6$ is
\be
g_{MN} = \begin{pmatrix} \eta_{\mu \nu} & 0 \\ 0 & -g_{\alpha \beta} \end{pmatrix} ~, 
\ee
where $\eta_{\mu \nu} = \mbox{diag}(1,-1,-1,-1)$ and $g_{\alpha \beta} = R^2 \mbox{diag}(1,\sin^2 \theta)$ are the metrics associated with $M^4$ and $S^2$, respectively.  The action in six-dimensional spacetime is then
\be
S_6 
= \int dx^4 d R^2 \Omega 
\biggl[ 
\bar{\Psi} i \Gamma^{\mu} D_{\mu} \Psi 
+ \bar{\Psi} i \Gamma^{a} e^{\alpha}_{a} D_{\alpha} \Psi 
- \frac{1}{4g^2} Tr[F_{MN}F^{MN}]
\biggr]
\ee
where $D_{M}$ ($M=0,1,2,3,\theta,\phi$) are covariant derivatives, $\Gamma^{\mu,a}$ are the Dirac gamma matrices in six dimensions, and $e^{\alpha}_a$ are the vielbeins on the two-sphere.  Explicitly,
\begin{equation}\begin{array}{lll}
D_{\mu} = \partial_{\mu} - iA_{\mu},
&
D_{\theta} = \partial_{\theta} -i A_{\theta},
&
D_{\phi} = \partial_{\phi} -i \frac{\Sigma_3}{2} \cos \theta -iA_{\phi}, \\
\Gamma_{\mu} = \gamma_{\mu} \otimes \mathbf{I}_2,
&
\Gamma_4 = \gamma_{5} \otimes \sigma_1,
&
\Gamma_5 = \gamma_{5} \otimes \sigma_2, \\
e^1_{\theta} = R,
&
e^2_{\phi} = R \sin \theta,
&
e^1_{\phi} = e^2_{\theta} = 0,
\end{array}\end{equation}
where $\sigma_i \, (i=1,2,3)$ are the Pauli matrices, $\mathbf{I}_d$ is the $d \times d$ identity matrix, and $\Sigma_3$ is defined as $\Sigma_3=\mathbf{I}_4 \otimes \sigma_3$.  The gauge field strength is $F_{MN}=\partial_M A_N -\partial_N A_M -i[A_M,A_N]$.  Note that the covariant derivative $D_{\phi}$ has a spin connection term $i \frac{\Sigma_3}{2} \cos \theta$ for fermions because of the nonzero curvature of the two-sphere.  This term generally induces a fermion mass in the four-dimensional effective action after integrating out the extra space.  This mass term, as we will see in Section~\ref{sec:fermions}, can be avoided by introducing a background gauge field $A^B_{\phi} \equiv {\tilde A}^B_\phi \sin\theta$ that corresponds to a Dirac monopole \cite{RandjbarDaemi:1982hi}
\be
\label{background}
{\tilde A}^B_{\phi} = - Q \frac{\cos \theta \mp 1}{\sin \theta} ~, \quad (-: 0 \leq \theta< \frac{\pi}{2} ~, \quad +: \frac{\pi}{2} \leq \theta \leq \pi)
\ee
where $Q$ is proportional to the generator of a U(1) subgroup of the original gauge group E$_6$.

%%%%%%%%%%%%%%%%%%%%%%%%%%%%%%%%%%%%%%%%%%%%%%%%%%%%%%%%%%%%%%%%%%%%%%%%%%%%%%%%%%%%%%%%
 
\subsection{Boundary conditions on the two-sphere orbifold \label{sec:bc}}

On the two-sphere orbifold, one can consider two parity operations $P_1: \, (\theta,\phi) \to (\pi-\theta,\phi)$ and $P_2: \, (\theta,\phi) \to (\pi-\theta,2\pi-\phi)$, which are related to each other by an azimuthal translation $\phi \to \phi+2\pi$.  We impose the following boundary conditions on both gauge and fermion fields under the two parity operations:
\bea
\label{boundary-condition1}
A_{\mu} (x,\pi-\theta,-\phi) 
&=& P_1 A_{\mu}(x,\theta,\phi) P_1 ~, \\
\label{boundary-condition2}
A_{\theta,\phi}(x,\pi-\theta,-\phi) 
&=& - P_1 A_{\theta,\phi}(x,\theta,\phi) P_1 ~, \\
\label{boundary-condition3}
\Psi (x,\pi-\theta,-\phi) 
&=& \pm \gamma_5 P_1 \Psi(x,\theta,\phi) ~, \\
\label{boundary-condition4}
A_{\mu} (x,\pi-\theta,2\pi-\phi) 
&=& P_2 A_{\mu}(x,\theta,\phi) P_2 ~, \\
\label{boundary-condition5}
A_{\theta,\phi}(x,\pi-\theta,2\pi-\phi) 
&=& - P_2 A_{\theta,\phi}(x,\theta,\phi) P_2 ~, \\
\label{boundary-condition6}
\Psi (x,\pi-\theta,2\pi-\phi) &=& \pm \gamma_5 P_2 \Psi(x,\theta,\phi) ~.
\eea
These boundary conditions are determined by requiring the invariance of the six -dimensional action under the transformation $(\theta,\phi) \rightarrow (\pi-\theta,-\phi)$.

The projection matrices $P_{1,2}$ act on the gauge group representation space and have eigenvalues $\pm 1$.  They assign different parities for different representation components.  For fermion boundary conditions, the sign in front of $\gamma_5$ can be either $+$ or $-$ since the fermions always appear in bilinear forms in the action.  The 4-dimensional action is then restricted by these parity assignments and our choice of the background gauge field.

%%%%%%%%%%%%%%%%%%%%%%%%%%%%%%%%%%%%%%%%%%%%%%%%%%%%%%%%%%%%%%%%%%%%%%%%%%%%%%%%%%%%%%%%

\subsection{Gauge group reduction \label{sec:group}}

We consider the following gauge group reduction
\bea
\label{group-red}
E_6 &\supset& SO(10) \times U(1)_Z \nonumber \\
&\supset& SU(5) \times U(1)_X \times U(1)_Z \nonumber \\
&\supset& SU(3) \times SU(2) \times U(1)_Y \times U(1)_X \times U(1)_Z ~.
\eea
The background gauge field in Eq.~(\ref{background}) is chosen to belong to the U(1)$_Z$ group.  This choice is needed in order to obtain chiral SM fermions in four dimensions to be discussed later.  There are two other symmetry reduction schemes.  One can prove that the results in those two schemes are effectively the same as the one considered here once we require the correct U(1) combinations for the hypercharge and the background field.

We then impose the parity assignments with respect to the fixed points, Eqs.~(\ref{boundary-condition1})-(\ref{boundary-condition6}).  The parity assignments for the fundamental representation of E$_6$ is chosen to be
\bea
\label{d27}
{\bf 27} &=& (1,2)(-3,-2,-2)^{(+,+)}+(1,2)(3,2,-2)^{(-,-)}+(1,2)(-3,3,1)^{(+,-)}  \nonumber \\ 
&& + (1,1)(6,-1,1)^{(+,+)}+(1,1)(0,-5,1)^{(-,-)}+(1,1)(0,0,4)^{(-,+)}  \nonumber \\
&& + (3,2)(1,-1,1)^{(-,+)}+(3,1)(-2,2,-2)^{(+,-)} + (\bar{3},1)(-4,-1,1)^{(+,+)} \nonumber \\
&& +(\bar{3},1)(2,3,1)^{(+,+)}+(\bar{3},1)(2,-2,2)^{(-,+)}, 
\eea
where, for example, $(+,-)$ means that the parities under $P_1$ and $P_2$ are (even,odd).  By the requirement of consistency, we find that the components of $A_{\mu}$ in the adjoint representation have the parities under $A_{\mu} \rightarrow P_1 A_{\mu} P_1$ $(P_2 A_{\mu} P_2)$ as follows:
\begin{eqnarray}
\label{d78}
{\bf 78}|_{A_{\mu}} &=& \underline{(8,1)(0,0,0)^{(+,+)}+(1,3)(0,0,0)^{(+,+)}} \nonumber \\
&& + \underline{(1,1)(0,0,0)^{(+,+)} +(1,1)(0,0,0)^{(+,+)}+(1,1)(0,0,0)^{(+,+)}}  \nonumber \\
&& + (3,2)(-5,0,0)^{(-,+)}+(\bar{3},2)(5,0,0)^{(-,+)}+   (3,2)(1,4,0)^{(+,-)}+ (\bar{3},2)(-1,-4,0)^{(+,-)}  \nonumber \\
&&  +(3,1)(4,-4,0)^{(-,-)}+(\bar{3},1)(-4,4,0)^{(-,-)}+(1,1)(-6,-4,0)^{(-,-)}+(1,1)(6,4,0)^{(-,-)} \nonumber \\
&& +  (3,2)(1,-1,-3)^{(+,+)}+  (\bar{3},2)(-1,1,3)^{(+,+)}+(3,1)(4,1,3)^{(-,+)}+(\bar{3},1)(-4,-1,-3)^{(-,+)}   \nonumber \\
&& + (3,1)(-2,-3,3)^{(+,-)}+ (\bar{3},1)(2,3,-3)^{(+,-)}+(1,2)(-3,3,-3)^{(-,-)}+(1,2)(3,-3,3)^{(-,-)}    \nonumber \\
&& +(1,1)(-6,1,3)^{(-,+)}+(1,1)(6,-1,-3)^{(-,+)}+(1,1)(0,-5,-3)^{(+,-)}+(1,1)(0,5,3)^{(+,-)},    \nonumber \\
\end{eqnarray}
where the underlined components correspond to the adjoint representations of SU(3) $\times$ SU(2) $\times$ U(1)$_Y$ $\times$ U(1)$_X$ $\times$ U(1)$_Z$, respectively.  We note that the components with parity $(+,+)$ can have massless zero modes in four dimensions.  Such components include the adjoint representations of SU(3) $\times$ SU(2) $\times$ U(1)$^3$, $(3,2)(1,-1,-3)$ and its conjugate.  The latter components seem problematic since they do not appear in the low-energy spectrum.  In fact, these components acquire masses due to the background field from the term proportional to $F_{\mu \phi} F^{\mu}_{\ \phi}$
\bea
&& Tr\left[
-\frac{1}{4} F_{\mu \nu}F^{\mu \nu} 
+ \frac{1}{2R^2 \sin^2 \theta} F_{\mu \phi}F^{\mu}_{\ \phi}
\right] \nonumber \\ 
&& \quad \rightarrow 
Tr\left[
-\frac{1}{4} (\partial_{\mu} A_{\nu}-\partial_{\nu} A_{\mu})(\partial^{\mu} A^{\nu}-\partial^{\nu} A^{\mu}) 
- \frac{1}{2R^2 \sin^2 \theta} [A_{\mu},A^B_{\phi}][A^{\mu},A^B_{\phi}]
\right] ~.
\eea
For the components of $A_{\mu}$ with nonzero U(1)$_Z$ charge, we have 
\begin{equation}
A_{\mu}^i Q_i+ A_{i\mu} Q^{i} \in A_{\mu} ~,
\end{equation}
where $Q_i \, (Q^i = Q_i^{\dagger})$ are generators corresponding to distinct components in Eq.~(\ref{d78}) that have nonzero U(1)$_Z$ charges, and $A_{i\mu} \, (A_{\mu}^i = A_{i \mu}^{\dagger})$ are the corresponding components of $A_{\mu}$.  We then find the term
\begin{eqnarray}
\frac{1}{\sin^2 \theta}Tr[[A_{\mu},A^B_\phi][A^{\mu},A^B_\phi]]
&=& \frac{(\cos \theta \mp 1)^2}{\sin^2 \theta}   
Tr[[A_{\mu}^i Q_i+A_{i \mu} Q^i,Q][ A^{i \mu} Q_i+A_{i}^{\mu} Q^i,Q]] \nonumber \\
&=& -2 |q|^2 \frac{(\cos \theta \mp 1)^2}{\sin^2 \theta} A^{i \mu} A_{i \mu} ~,
\end{eqnarray}
where $q$ is the $Q$ charge of the relevant component.  Use of the facts that $A_{\phi}^B$ belongs to U(1)$_Z$ and that $Tr[Q_i Q^i]=2$ has been made in the above equation.  A mass is thus associated with the lowest modes of those components of $A_{\mu}$ with nonzero U(1)$_Z$ charges:
\bea
&& \int d \Omega 
Tr\left.\left[
-\frac{1}{4}(\partial_{\mu} A_{\nu}-\partial_{\nu} A_{\mu})(\partial^{\mu} A^{\nu}-\partial^{\nu} A^{\mu})  - \frac{1}{2R^2 \sin^2 \theta} [A_{\mu},A_B][A^{\mu},A_B]
\right] \right|_{\rm lowest} \nonumber \\
&& \quad \rightarrow 
-\frac{1}{2} 
\left[ \partial_{\mu} A_{i \nu}(x) - \partial_{\nu} A_{i \mu}(x) \right]
\left[ \partial^{\mu} A^{i \nu}(x) - \partial^{\nu} A^{i \mu}(x) \right]
+ m^2_B A_{\mu}^i(x) A^{i\mu}(x) ~,
\eea
where the subscript `lowest' means that only the lowest KK modes are kept.  
Here the lowest KK modes of $A_{\mu}$ correspond to the term $A_{\mu}(x)/\sqrt{4 \pi}$ in the KK expansion.  In summary, any representation of $A_\mu$ carrying a nonzero U(1)$_Z$ charge acquires a mass $m_B$ from the background field contribution after one integrates over the extra spatial coordinates.  More explicitly,
\be
\label{eq:nonSMgaugeMass}
m^2_B = \frac{|q|^2}{4 \pi R^2} 
\int d \Omega \frac{(\cos \theta \mp 1)^2}{\sin^2 \theta} 
\simeq 0.39 \frac{ |q|^2}{R^2} 
\ee
for the zero mode.  Therefore, the $(3,2)(1,-1,-3)$ representation and its conjugate are elevated in mass to disappear from the low-energy spectrum.  In the end, the correct symmetry reduction is achieved since only the components of 4-dimensional gauge field $A_{\mu}$ in the adjoint representation of SU(3) $\times$ SU(2) $\times$ U(1)$_Y$ $\times$ U(1)$_X$ $\times$ U(1)$_Z$ are allowed to have zero modes.  A general discussion about the KK mode masses of $A_{\mu}$ will be given in Section~\ref{KKmass}.

\subsection{Scalar field contents in four dimensions \label{sec:scalar}}

The scalar contents in four dimensions are obtained from the extra-dimensional components of the gauge field $\{ A_{\theta}, A_{\phi} \}$ after integrating out the extra spatial coordinates.  The kinetic term and potential term of $\{A_{\theta}, A_{\phi} \}$ are obtained from the gauge sector containing these components
\begin{eqnarray}
\label{action-scalar}
S_{\rm scalar}
&=& \int dx^4 d \Omega \Bigl(  \frac{1}{2 g^2}  Tr[ F_{\mu \theta} F^{\mu}_{\  \theta} ] + \frac{1}{2 g^2 \sin^2 \theta}  Tr[ F_{\mu \phi} F^{\mu}_{\  \phi} ] \nonumber \\
&& \qquad \qquad  -\frac{1}{2 g^2 R^2 \sin^2 \theta}  Tr[ F_{\theta \phi} F_{\theta \phi} ] \Bigr) \nonumber \\
&\rightarrow& \int dx^4 d \Omega \Bigl( \frac{1}{2 g^2} Tr[(\partial_{\mu} A_{\theta}-i[A_{\mu},A_{\theta}])^2] 
+ \frac{1}{2 g^2} Tr[(\partial_{\mu} A_{\theta}-i[A_{\mu},\tilde{A}_{\phi}])^2 ] \nonumber \\
&& \qquad \qquad -\frac{1}{2 g^2 R^2} Tr \biggl[  \biggl( \frac{1}{\sin \theta} \partial_{\theta} (\sin \theta \tilde{A}_{\phi} + \sin \theta \tilde{A}^B_{\phi}) -\frac{1}{\sin \theta} \partial_{\phi} A_{\theta} 
- i[A_{\theta},\tilde{A}_{\phi}+\tilde{A}^B_{\phi}] \biggr)^2 \biggr] ~, \nonumber \\
\end{eqnarray}
where we have taken $A_{\phi} = \tilde{A}_{\phi} \sin \theta + \tilde{A}_{\phi}^B \sin \theta$.  In the second step indicated by the arrow in Eq.~(\ref{action-scalar}), we have omitted terms which do not involve $A_{\theta}$ and $\tilde{A}_{\phi}$ from the right-hand side of the first equality.  It is known that one generally cannot obtain massless modes for physical scalar components in four dimensions \cite{Maru:2006, Dohi:2010vc}.  One can see this by noting that the eigenfunction of the operator $\frac{1}{ \sin \theta } \partial_{\theta} \sin \theta$ with zero eigenvalue is not normalizable \cite{Maru:2006}.  In other words, these fields have only KK modes.  However, an interesting feature is that it is possible to obtain a negative squared mass when taking into account the interactions between the background gauge field $\tilde{A}_{\phi}^B$ and $\{A_{\theta}, \tilde{A}_{\phi} \}$.  This happens when the component carries a nonzero U(1)$_Z$ charge, as the background gauge field belongs to U(1)$_Z$.  In this case, the $(\ell=1,m=1)$ modes of these real scalar components are found to have a negative squared mass in four dimensions.  They can be identified as the Higgs fields once they are shown to belong to the correct representation under the SM gauge group.  Here the numbers $(\ell,m)$ are the angular momentum quantum number on $S^2/Z_2$, and each KK mode is characterized by these numbers.  One can show that the $(\ell=1,m=0)$ mode has a positive squared mass and is not considered as the Higgs field.  A discussion of the KK masses with general $(\ell,m)$ will be given in Section~\ref{KKmass}~.

With the parity assignments with respect to the fixed points, Eqs.~(\ref{boundary-condition2}) and (\ref{boundary-condition5}), we have for the $A_{\theta}$ and $A_{\phi}$ fields
\bea
\label{78scalar}
{\bf 78}|_{A_{\theta,\phi}} & = &  (8,1)(0,0,0)^{(-,-)}+(1,3)(0,0,0)^{(-,-)} \nonumber \\
&& +(1,1)(0,0,0)^{(-,-)} +(1,1)(0,0,0)^{(-,-)}+(1,1)(0,0,0)^{(-,-)}  \nonumber \\
&& + (3,2)(-5,0,0)^{(+,-)}+(\bar{3},2)(5,0,0)^{(+,-)}+   (3,2)(1,4,0)^{(-,+)}+ (\bar{3},2)(-1,-4,0)^{(-,+)}  \nonumber \\
&&  +(3,1)(4,-4,0)^{(+,+)}+(\bar{3},1)(-4,4,0)^{(+,+)}+(1,1)(-6,-4,0)^{(+,+)}+(1,1)(6,4,0)^{(+,+)} \nonumber \\
&& +  (3,2)(1,-1,-3)^{(-,-)}+  (\bar{3},2)(-1,1,3)^{(-,-)}+(3,1)(4,1,3)^{(+,-)}+(\bar{3},1)(-4,-1,-3)^{(+,-)}   \nonumber \\
&& + (3,1)(-2,-3,3)^{(-,+)}+ (\bar{3},1)(2,3,-3)^{(-,+)}+(1,2)(-3,3,-3)^{(+,+)}+(1,2)(3,-3,3)^{(+,+)}    \nonumber \\
&& +(1,1)(-6,1,3)^{(+,-)}+(1,1)(6,-1,-3)^{(+,-)}+(1,1)(0,-5,-3)^{(-,+)}+(1,1)(0,5,3)^{(-,+)} ~. \nonumber \\
\eea
Components with $(+,-)$ or $(-,+)$ parity do not have KK modes since they are odd under $\phi \rightarrow \phi+2\pi$ and the KK modes of gauge field are specified by integer angular momentum quantum numbers $\ell$ and $m$ on the two-sphere.  We then concentrate on the components which have either $(+,+)$ or $(-,-)$ parity and nonzero U(1)$_Z$ charges as the candidate for the Higgs field.  These include $\{ (1,2)(3,-3,3) + {\rm h.c.} \}$ and $\{(3,2)(1,-1,-3) + {\rm h.c.} \}$ with parities $(+,+)$ and $(-,-)$, respectively.  The representations $(1,2)(-3,3,-3)$ and $(1,2)(3,-3,3)$ have the correct quantum numbers for the SM Higgs doublet.  Therefore, we identify the $(1,1)$ mode of these components as the SM Higgs fields in four dimensions.

\subsection{Chiral fermions in four dimensions \label{sec:fermions}}

We introduce fermions as the Weyl spinor fields of the six-dimensional Lorentz group SO(1,5).  They can be written in terms of the SO(1,3) Weyl spinors as
\bea
\label{chiralR}
\Psi_+ = \begin{pmatrix} \psi_R \\ \psi_L \end{pmatrix} ~, \\
\label{chiralL} 
\Psi_- = \begin{pmatrix} \psi_L \\ \psi_R \end{pmatrix} ~.
\eea
In general, fermions on the two-sphere do not have massless KK modes because of the positive curvature of the two-sphere.  The massless modes can be obtained by incorporating the background gauge field (\ref{background}) though, for it can cancel the contribution from the positive curvature.  In this case, the condition for obtaining a massless fermion mode is
\be
\label{massless-condition}
Q \Psi = \pm \frac{1}{2} \Psi ~,
\ee
where $Q$ comes from the background gauge field and is proportional to the U(1)$_Z$ generator \cite{RandjbarDaemi:1982hi,Maru:2009wu,Dohi:2010vc}.  
We observe that the upper [lower] component on the RHS of Eq.~(\ref{chiralR}) [(\ref{chiralL})] has a massless mode for the $+$ $(-)$ sign on the RHS of Eq.~(\ref{massless-condition}).

In our model, we choose the fermions as the Weyl fermions $\Psi_-$ belonging to the {\bf 27} representation of E$_6$.  The {\bf 27} representation is decomposed as in Eq.~(\ref{d27}) under the group reduction, Eq.~(\ref{group-red}).  In this decomposition, we find that our choice of the background gauge field of U(1)$_Z$ is suitable for obtaining massless fermions since all such components have U(1)$_Z$ charge 1.  In the fundemantal representation, the U(1)$_Z$ generator is
\bea
Q_Z = \frac{1}{6} \mbox{diag}
(-2,-2,-2,-2,1,1,1,1,4,1,1,1,1,1,1,-2,-2,-2,1,1,1,1,1,1,-2,-2,-2) ~,
\nonumber \\
\eea
according to the decomposition Eq.~(\ref{d27}).  By identifying $Q=3Q_Z$, we readily obtain the condition
\begin{equation}
Q \Psi_- = \frac{1}{2} \Psi_-.
\end{equation}
Therefore, the chiral fermions $\psi_L$ in four dimensions have zero modes.

Next, we consider the parity assignments for the fermions with respect to the fixed points of $S^2/Z_2$.  The boundary conditions are given by Eqs.~(\ref{boundary-condition3}) and (\ref{boundary-condition6}).  It turns out that four ${\bf 27}$ fermion copies with different boundary conditions are needed in order to obtain an entire generation of massless SM fermions.  They are denoted by $\Psi^{(1,2,3,4)}$ with the following parity assignments
\bea
\Psi_{\pm}^{(i)} (x,\pi-\theta,-\phi) 
&=& \xi \gamma_5 P_1 \Psi_{\pm}^{(i)}(x,\theta,\phi) ~, \\
\Psi_{\pm}^{(i)} (x,\pi-\theta,2\pi-\phi) 
&=& \eta \gamma_5 P_2 \Psi_{\pm}^{(i)}(x,\theta,\phi) ~,
\eea
where $\gamma_5$ is the chirality operator, and $(\xi,\eta) = (+,+)$, $(-,-)$, $(+,-)$ and $(-,+)$ for $i = 1,2,3,4$, respectively.  From these fermions we find that $\psi_{1,2,3,4}$ have the parity assignments
\bea
{\bf 27}_{\psi_L^{(1)}} &=& (1,2)(-3,-2,-2)^{(-,-)}+(1,2)(3,2,-2)^{(+,+)}+(1,2)(-3,3,1)^{(-,+)}  \nonumber \\ 
&& + (1,1)(6,-1,1)^{(-,-)}+ \underline{(1,1)(0,-5,1)^{(+,+)}}+(1,1)(0,0,4)^{(+,-)}  \nonumber \\
&& + (3,2)(1,-1,1)^{(+,-)}+(3,1)(-2,2,-2)^{(-,+)} + (\bar{3},1)(-4,-1,1)^{(-,-)} \nonumber \\
&& +(\bar{3},1)(2,3,1)^{(-,-)}+(\bar{3},1)(2,-2,2)^{(+,-)} \\
{\bf 27}_{\psi_L^{(2)}} &=& (1,2)(-3,-2,-2)^{(+,+)}+(1,2)(3,2,-2)^{(-,-)}+(1,2)(-3,3,1)^{(+,-)}  \nonumber \\ 
&& + \underline{(1,1)(6,-1,1)^{(+,+)} } + (1,1)(0,-5,1)^{(-,-)}+(1,1)(0,0,4)^{(-,+)}  \nonumber \\
&& + (3,2)(1,-1,1)^{(-,+)}+(3,1)(-2,2,-2)^{(+,-)} + \underline{ (\bar{3},1)(-4,-1,1)^{(+,+)} } \nonumber \\
&& + \underline{(\bar{3},1)(2,3,1)^{(+,+)} }+(\bar{3},1)(2,-2,2)^{(-,+)} \\
{\bf 27}_{\psi_L^{(3)}} &=& (1,2)(-3,-2,-2)^{(-,+)}+(1,2)(3,2,-2)^{(+,-)}+(1,2)(-3,3,1)^{(-,-)}  \nonumber \\ 
&& + (1,1)(6,-1,1)^{(-,+)}+(1,1)(0,-5,1)^{(+,-)}+(1,1)(0,0,4)^{(+,+)}  \nonumber \\
&& + \underline{ (3,2)(1,-1,1)^{(+,+)} } +(3,1)(-2,2,-2)^{(-,-)} + (\bar{3},1)(-4,-1,1)^{(-,+)} \nonumber \\
&& +(\bar{3},1)(2,3,1)^{(-,+)}+(\bar{3},1)(2,-2,2)^{(+,+)} \\
{\bf 27}_{\psi_L^{(4)}} &=& (1,2)(-3,-2,-2)^{(+,-)}+(1,2)(3,2,-2)^{(-,+)}+ \underline{ (1,2)(-3,3,1)^{(+,+)} }  \nonumber \\ 
&& + (1,1)(6,-1,1)^{(+,-)}+(1,1)(0,-5,1)^{(-,+)}+(1,1)(0,0,4)^{(-,-)}  \nonumber \\
&& + (3,2)(1,-1,1)^{(-,-)}+(3,1)(-2,2,-2)^{(+,+)} + (\bar{3},1)(-4,-1,1)^{(+,-)} \nonumber \\
&& +(\bar{3},1)(2,3,1)^{(+,-)}+(\bar{3},1)(2,-2,2)^{(-,-)} ~, 
\eea
where the underlined components have even parities and U(1)$_Z$ charge 1.  
One can readily identify one generation of SM fermions, including a right-handed neutrino, as the zero modes of these components.

A long-standing problem in the gauge-Higgs unification framework is the Yukawa
couplings of the Higgs boson to the matter fields.  This is because the
couplings here all arise from gauge interactions.  It is therefore extremely
difficult to derive the observed rich fermion mass spectrum purely from the gauge
coupling.  In order to have flavor-dependent Yukawa couplings, one promising
solution is to consider SM matter fields localized at orbifold fixed points and
make use of nonlocal interactions with Wilson lines \cite{Csaki:2002ur}.

\section{Higgs potential \label{sec:higgs}}

\subsection{Higgs sector \label{sec:Higgssector}}

The Lagrangian for the Higgs sector is derived from the gauge sector that contains extra-dimensional components of the gauge field $\{A_{\theta}, \tilde{A}_{\phi} \}$, as given in Eq.~(\ref{action-scalar}), by considering the lowest KK modes of them.  The kinetic term and potential term are, respectively,
\bea
L_{K}
&=& \frac{1}{2 g^2} \int d \Omega 
\left.
\Bigl( Tr[(\partial_{\mu} A_{\theta}-i[A_{\mu},A_{\theta}])^2] 
+ Tr[(\partial_{\mu} A_{\theta}-i[A_{\mu},\tilde{A}_{\phi}])^2 ]  \Bigr)
\right|_{\textrm{lowest}} ~, \\
V
&=& \frac{1}{2 g^2 R^2} \int d \Omega 
\left.
Tr \biggl[  \biggl( \frac{1}{\sin \theta} \partial_{\theta} (\sin \theta \tilde{A}_{\phi} 
+ \sin \theta \tilde{A}^B_{\phi}) -\frac{1}{\sin \theta} \partial_{\phi} A_{\theta} 
- i[A_{\theta},\tilde{A}_{\phi}+\tilde{A}^B_{\phi}] \biggr)^2 \biggr]
\right|_{\textrm{lowest}} ~. \nonumber \\
\eea
Consider the $(1,1)$ mode of the $ \{ (1,2)(3,-3,3) + {\rm h.c.} \}$ representation in Eq.~(\ref{78scalar}) as argued in the previous section.  The gauge fields are given by the following KK expansions
\bea
\label{expansion1}
A_{\theta} &=& - \frac{1}{\sqrt{2}} [ \Phi_1(x)  \partial_{\theta} Y_{11}^-( \theta, \phi) + \Phi_2(x) \frac{1}{\sin \theta} \partial_{\phi} Y_{11}^-( \theta, \phi) ] + \cdots ~, \\
\label{expansion2}
\tilde{A}_{\phi} &=& \frac{1}{\sqrt{2}}[ \Phi_2(x)  \partial_{\theta} Y_{11}^-( \theta,\phi)-\Phi_1(x) \frac{1}{\sin \theta}  \partial_{\phi} Y_{11}^-( \theta,\phi)] + \cdots ~,
\eea
where $\cdots$ represents higher KK mode terms \cite{Maru:2009wu}.   The function $Y_{11}^- = -1/\sqrt{2} [Y_{11}+Y_{1-1}]$ is odd under $(\theta,\phi) \rightarrow (\pi/2-\theta,-\phi)$ .  We will discuss their higher KK modes and masses in the existence of the background gauge field in Section~\ref{KKmass}.  With Eqs.~(\ref{expansion1}) and (\ref{expansion2}), the kinetic term becomes
\bea
L_{K}(x) = \frac{1}{2 g^2}  \Bigl( Tr[D_{\mu} \Phi_1(x) D^{\mu} \Phi_1(x)] + Tr[D_{\mu} \Phi_2(x) D^{\mu} \Phi_2(x)]  \Bigr)  ,
\eea
where $D_{\mu} \Phi_{1,2} = \partial_{\mu} \Phi_{1,2} -i[A_{\mu},\Phi_{1,2}]$ is the covariant derivative acting on $\Phi_{1,2}$.  The potential term, on the other hand, is
\bea
V 
&=& \frac{1}{2 g^2 R^2} \int d \Omega Tr \biggl[ \biggl( -\sqrt{2} Y_{11}^- \Phi_2(x) +  Q  
+ \frac{i}{2}  [\Phi_1(x), \Phi_2(x)] \{ \partial_{\theta} Y_{11}^- \partial_{\theta} Y_{11}^- + \frac{1}{\sin^2 \theta} \partial_{\phi} Y_{11}^- \partial_{\phi} Y_{11}^- \} \nonumber \\
&& \qquad \qquad 
+\frac{ i}{\sqrt{2}} [\Phi_1(x), \tilde{A}^B_{\phi}] \partial_{\theta} Y_{11}^- +\frac{ i}{\sqrt{2}} [\Phi_2(x), \tilde{A}^B_{\phi}] \frac{1}{\sin \theta} \partial_{\phi} Y_{11}^-  \biggr)^2 \biggr] ~,
\eea
where $\partial_{\theta} (\sin \theta \tilde{A}_{\phi}^B) = Q \cos \theta$ from Eq.~(\ref{background}) is used.  Expanding the square in the trace, we get
\bea
\label{potential}
V &=& \frac{1}{2 g^2 R^2} \int  d \Omega Tr
\biggl[ 2 (Y_{11}^+)^2 \Phi_2^2(x) + Q^2 
- \frac{1}{4} [\Phi_1(x),\Phi_2(x)]^2 \left( \partial_{\theta} Y_{11}^- \partial_{\theta} Y_{11}^- + \frac{1}{\sin^2 \theta} \partial_{\phi} Y_{11}^- \partial_{\phi} Y_{11}^- \right)^2  \nonumber \\
&& \qquad \qquad \qquad 
-\frac{1}{2} [\Phi_1(x),\tilde{A}^B_{\phi}]^2  (\partial_{\theta} Y_{11}^- )^2 -\frac{1}{2} [\Phi_2(x),\tilde{A}^B_{\phi}]^2  \left( \frac{1}{\sin \theta} \partial_{\phi} Y_{11}^- \right)^2 \nonumber \\ 
&& \qquad \qquad \qquad 
-2 i \Phi_2(x) [\Phi_1(x), \tilde{A}_{\phi}^B] 
Y_{11}^- \partial_{\theta} Y_{11}^-
- [\Phi_1(x),\tilde{A}_{\phi}^B] [\Phi_2(x),\tilde{A}_{\phi}^B] 
\partial_{\theta} Y_{11}^- \frac{1}{\sin \theta} \partial_{\phi} Y_{11}^- \nonumber \\
&& \qquad \qquad \qquad 
+ i Q [\Phi_1(x), \Phi_2(x)] \left( \partial_{\theta} Y_{11}^- \partial_{\theta} Y_{11}^- + \frac{1}{\sin^2 \theta} \partial_{\phi} Y_{11}^- \partial_{\phi} Y_{11}^- \right) 
~\biggr] ~,
\eea
where terms that vanish after the $d\Omega$ integration are directly omitted.
In the end, the potential is simplified to
\bea
V = \frac{1}{2 g^2 R^2} Tr \biggl[ 2 \Phi_2^2(x) + 4 \pi Q^2 - \frac{3}{10 \pi} [\Phi_1(x),\Phi_2(x)]^2 +  \frac{5i}{2} Q [\Phi_1(x), \Phi_2(x)]  \nonumber \\
+ \mu_1 [Q, \Phi_1(x)]^2 +  \mu_2 [Q, \Phi_2(x)]^2  \biggr] ~, 
\eea
where use of $\tilde{A}_{\phi}^B = -Q (\cos \theta \mp 1) / \sin \theta$ has been made and $\mu_1 = 1-\frac{3}{2} \ln 2$ and $\mu_2 = \frac{3}{4}(1-2\ln2)$.

We now take the following linear combination of $\Phi_1$ and $\Phi_2$ to form a complex Higgs doublet, 
\begin{eqnarray}
\label{okikae1}
\Phi(x) &=& \frac{1}{\sqrt{2}} (\Phi_1(x)+i\Phi_2(x)) ~, \\
\label{okikae2}
\Phi(x)^{\dagger} &=& \frac{1}{\sqrt{2}} (\Phi_1(x)-i\Phi_2(x)) ~.
\end{eqnarray}
It is straightforward to see that
\begin{eqnarray}
[\Phi_1(x),\Phi_2(x)] = i [\Phi(x), \Phi^{\dagger}(x)] ~.
\end{eqnarray}
The kinetic term and the Higgs potential now become 
\bea
\label{kinetic-t}
L_{K} &=& \frac{1}{g^2} Tr[D_{\mu} \Phi^{\dagger}(x) D^{\mu} \Phi(x) ] ~, \\
\label{potential-t}
V &=& \frac{1}{2 g^2 R^2} Tr \biggl[ 2 \Phi_2^2(x) + 4 \pi Q^2 + \frac{3}{10 \pi} [\Phi(x),\Phi^{\dagger}(x)]^2 - \frac{5}{2}  Q [\Phi(x), \Phi^{\dagger}(x)] \nonumber \\
&& \qquad + \mu_1[Q, \Phi_1(x)]^2 + \mu_2[Q, \Phi_2(x)]^2  \biggr] ~. 
\eea

To further simplify the above expressions, we need to find out the algebra of the gauge group generators.  Note that the E$_6$ generators are chosen according to the decomposition of the adjoint representation given in Eq.~(\ref{d78})
\bea
&&\{ Q_i, Q_{\alpha}, Q_Y, Q_X, Q_Z, \nonumber \\
&& \quad Q_{ax (-5,0,0)}, Q^{ax(5,0,0)}, Q_{ax(1,4,0)}, Q^{ax(-1,-4,0)}, \nonumber \\
&& \quad Q_{a(4,-4,0)}, Q^{a(-4,4,0)}, Q_{(-6,-4,0)}, Q_{(6,4,0)}, \nonumber \\
&& \quad Q_{ax(1,-1,-3)}, Q^{ax(-1,1,3)}, Q_{a(4,1,3)}, Q^{a(-4,-1,-3)}, \nonumber \\
&& \quad Q_{a(-2,-3,3)}, Q^{a(2,3,-3)}, Q_{x(3,-3,3)}, Q^{x(-3,3,-3)}, \nonumber \\
&& \quad Q_{(-6,1,3)}, Q_{(6,-1,-3)}, Q_{(0,-5,-3)},Q_{(0,5,3)}  \} ~,
\eea
where the generators are listed in the corresponding order of the terms in Eq.~(\ref{d78}) and the indices
\bea
\label{generators}
&& i=1,...,8: \textrm{SU(3) adjoint representation index} \Rightarrow Q_i:  \textrm{SU(3) generators} ~, \\
&& \alpha=1,2,3: \textrm{SU(2) adjoint representation index} \Rightarrow Q_{\alpha}: \textrm{SU(2) generators} ~, \\
&& Q_{X,Y,Z}: \textrm{$U(1)_{X,Y,Z}$ generators} ~, \\
&& x=1,2: \textrm{SU(2) doublet index} ~, \\
&& a=1,2,3: \textrm{SU(3) color index} ~.
\eea
Here we take the standard normalization for generators, $Tr[Q Q^{\dagger}] = 2$. 
The Higgs fields are in the representations of $(1,2)(3,-3,3)$ and $(1,2)(-3,3,-3)$.  We write
\begin{equation}
\label{Higgs}
\Phi(x) = \phi^{x} Q_{x(3,-3,3)} \quad (\Phi^{\dagger}(x) = \phi_x Q^{x(-3,3,-3)}) ~.
\end{equation}
Likewise, the gauge field $A_{\mu}(x)$ in terms of the $Q$'s in Eq.~(\ref{generators}) is
\be
\label{gauge}
 A_{\mu}(x) = A_{\mu}^i Q_i+A_{\mu}^{\alpha} Q_{\alpha}+B_{\mu} Q_Y+C_{\mu} Q_X+E_{\mu} Q_Z.
\ee
The commutation relations between the generators $Q_{\alpha}$, $Q_{X,Y,Z}$, $Q_{x(3,-3,3)}$ and $Q^{x(-3,3,-3)}$ are summarized in Table.~\ref{commutators}.

\begin{center}
\begin{table}
\begin{tabular}{lll}
\hline\hline
& \multicolumn{2}{c}{
$\left[ Q_{x(3,-3,3)},Q^{y(-3,3,-3)} \right] 
= \frac{1}{2} \delta_x^y Q_Z-\frac{1}{2} \sqrt{\frac{3}{5}} \delta_x^y Q_X
+ \frac{1}{\sqrt{10}} \delta_x^y Q_Y + \frac{1}{\sqrt{6}} (\sigma_{\alpha})^y_x Q_{\alpha}$}  \\
&
$\left[ Q_{\alpha},Q_{x(3,-3,3)} \right] = \frac{1}{\sqrt{6}} (\sigma_{\alpha} )^y_x Q_{y(3,-3,3)}$ \qquad \qquad & 
$\left[ Q_{\alpha},Q^{ x(-3,3,-3) } \right] = - \frac{1}{\sqrt{6}} (\sigma_{\alpha}^* )^y_x Q^{y(-3,3,-3) }$  \\ 
&
$\left[ Q_{x(3,-3,3)},Q_{y(3,-3,3)} \right] = 0$ &
$\left[ Q_Z, Q_{x(3,-3,3)} \right] = \frac{1}{2} Q_{x(3,-3,3)}$  \\
&
$\left[ Q_X, Q_{x(3,-3,3)} \right] = -\frac{1}{2} \sqrt{\frac{3}{5}} Q_{x(3,-3,3)}$ &
$\left[ Q_Y, Q_{x(3,-3,3)} \right] = \frac{1}{\sqrt{10}} Q_{x(3,-3,3)}$ \\
\hline\hline 
\end{tabular}
\caption{Commutation relations of $Q_{\alpha}$, $Q_{X,Y,Z}$, $Q_{x(3,-3,3)}$ and $Q^{x(-3,3,-3)}$, where $\sigma_i$ are the Pauli matrices.}
\label{commutators}
\end{table}
\end{center}

Finally, we obtain the Lagrangian associated with the Higgs field by applying Eqs.~(\ref{Higgs}, \ref{gauge}) to Eqs.~(\ref{kinetic-t}, \ref{potential-t}) and carrying out the trace.  Furthermore, to obtain the canonical form of kinetic terms, the Higgs field, the gauge field, and the gauge coupling need to be rescaled in the following way:
\bea
\label{notation}
&& \phi \rightarrow \frac{g}{\sqrt{2}} \phi \\
&& A_{\mu} \rightarrow \frac{g}{R}A_{\mu} \\
&& \frac{g}{\sqrt{6 \pi R^2}} = g_2 ~,
\eea
where $g_2$ denotes the SU(2) gauge coupling.  The Higgs sector is then given by
\be
{\cal L}_{\rm Higgs} = |D_{\mu} \phi|^2 - V(\phi)
\ee
where
\bea
\label{cova}
D_{\mu} \phi &=& 
\left[ \partial_{\mu} + i g_2 \frac{\sigma_{\alpha}}{2}  A_{\alpha \mu}
+ ig \frac{1}{\sqrt{40 \pi R^2}} B_{\mu} 
- ig \frac{1}{2 } \sqrt{\frac{3}{20 \pi R^2}} C_{\mu}
+ i g \frac{1}{2 \sqrt{4 \pi R^2}} E_{\mu} \right] \phi ~,  \\
\label{H-potential}
V &=&  -\frac{\chi}{8 R^2} \phi^{\dagger} \phi + \frac{3 g^2}{40 \pi R^2} \left(\phi^{\dagger} \phi \right)^2 ~,
\eea
where $\chi=7+9\mu_1+9\mu_2$.
We have omitted the constant term in the Higgs potential.  Comparing the potential derived above with the standard form $\mu^2\phi^\dagger\phi + \lambda (\phi^\dagger\phi)^2$ in the SM, we see that the model has a tree-level $\mu^2$ term that is negative and proportional to $R^{-2}$.  Moreover, the quartic coupling $\lambda = 3 g^2 / (40 \pi R^2)$ is related to the six-dimensional gauge coupling $g$ and grants perturbative calculations because it is about $0.16$, using the value of $R$ to be extracted in the next section.  Therefore, the order parameter in this model is controlled by a single parameter $R$, the compactification scale.

In fact, the $(1,1)$ mode of the $\{(3,2)(1,-1,-3) + {\rm h.c.}\}$ representation also has a negative squared mass term because it has the same $Q_z$ charge as the $\{(1,2)(3,-3,3) + {\rm h.c.}\}$ representation.  Therefore, it would induce not only electroweak symmetry breaking but also color symmetry breaking.  This undesirable feature can be cured by adding brane terms
\bea
\frac{\alpha}{R^2\sin^2\theta}
\left( F_{\theta\phi}^a F^{a\theta\phi} \right)^2
\delta\left( \theta-\frac{\pi}2 \right)
\left[
\delta(\phi)
+ \delta(\phi-\pi)
\right] ~,
\eea
where $a$ denotes the group index of the $\{(3,2)(1,-1,-3) + {\rm h.c.}\}$ representation.  These brane terms preserve the $Z'_2$ symmetry which corresponds to the symmetry under the transformation $(\phi \rightarrow \phi+\pi)$.  With an appropriate choice of the dimensionless constant $\alpha$, the squared mass of the $(1,1)$ can be lifted to become positive and sufficiently large.

\subsection{Spontaneous symmetry breaking and Higgs mass}

Due to a negative mass term, the Higgs potential in Eq.~(\ref{H-potential}) can induce the spontaneous symmetry breakdown: SU(2) $\times$ U(1)$_Y$ $\rightarrow$ U(1)$_{\rm EM}$ in the SM.  The Higgs field acquires a vaccum expectation value (VEV)
\bea
\langle \phi \rangle =
\frac{1}{\sqrt{2}} \begin{pmatrix} 0  \\ v \end{pmatrix} \mbox{ with }
v = \sqrt{\frac{5 \pi \chi}{3}} \frac{1}{g} 
\simeq \frac{4.6}{g} ~.
\eea
One immedialtey finds that the $W$ boson mass
\begin{equation}
m_W = \frac{g_2}{2} v = \frac{1}{6}\sqrt{ \frac{5 \chi}{2} } \frac{1}{R} 
\simeq \frac{0.53}{R},
\end{equation}
from which the compactification scale $R^{-1} \simeq 152$ GeV is inferred.  Moreover, the Higgs boson mass at the tree level is
\begin{equation}
m_H = \sqrt{\frac{3}{20 \pi}} \frac{g v}{R} = 3 \sqrt{\frac{2}{5}} m_W 
= \frac{\sqrt{\chi}}{2}  \frac{1}{R} ~,
\end{equation}
which is about $152$ GeV, numerically very close to the compactification scale.  Since the hypercharge of the Higgs field is $1/2$, the U(1)$_Y$ gauge coupling is derived from Eq.~(\ref{cova}) as
\begin{equation}
g_Y = \frac{g}{\sqrt{10 \pi R^2}} ~.
\end{equation}
The Weinberg angle is thus given by 
\begin{eqnarray}
\sin^2 \theta_W = \frac{g_Y^2}{g_2^2+g_Y^2}
= \frac{3}{8} ~,
\end{eqnarray}
and the $Z$ boson mass
\begin{eqnarray}
m_Z = \frac{m_W}{\cos \theta_W} = m_W \sqrt{\frac{8}{5}} ~,
\end{eqnarray}
both at the tree level.  These relations are the same as the SU(5) GUT at the unification scale.  This is not surprising because this part only depends on the group structure.

%%%%%%%%%%%%%%%%%%%%%%%%%%%%%%%%%%%%%%%%%%%%%%%%%%%%%%%%%%%%%%%%%%%%%%%%%%%%%%%%%%%%%%%%%

%%%%%%%%%%%%%%%%%%%%%%%%%%%%%%%%%%%%%%%%%%%%%%%%%%%%%%%%%%%%%%%%%%%%%%%%%%%%%%%%%%%%%%%%%

\section{KK mode spectrum of each field \label{KKmass}}

In this section, we compute the KK mass spectra of both fermion and gauge fields in the existence of the background gauge field.  The masses are basically conrtrolled by the compactification radius $R$ of the two-sphere.  They receive two kinds of contributions: one arising from the angular momentum in the $S^2$ space, and the other coming from the interactions with the background field.

\subsection{KK masses of fermions}

The KK masses for fermions have been given in Refs.~\cite{RandjbarDaemi:1982hi, Maru:2009wu, Dohi:2010vc}.  We give them in terms of our notation here:
\bea
\label{KKmass-fermion}
M_{\ell m}^{KK}(\psi_L) = \frac{1}{R} \sqrt{\ell(\ell+1)-\frac{4q^2-1}{4} } ~,
\eea
where $q$ is proportional to the U(1)$_Z$ charge of fermion and determined by the action of $Q=3Q_Z$ on fermions as $Q \Psi = q \Psi = 3q_Z \Psi$.   Note that the mass does not depend on the quantum number $m$.  The lightest KK mass, corresponding to $\ell = 1$ and $q_Z = 1/6$, is about 214 GeV at the tree level.
The range of $\ell$ is 
\be
\frac{2q \pm 1}{2} \leq \ell \qquad  (+: \ \textrm{for} \ \psi_{R(L)} \ \textrm{in} \ \Psi_{+(-)},  \quad - : \ \textrm{for} \ \psi_{L(R)} \ \textrm{in} \ \Psi_{-(+)} ) ~. 
\ee
We thus can have zero mode for $Q \Psi = \pm \frac{1}{2} \Psi$, where this condition is given in Eq.~(\ref{massless-condition}).  

\subsection{KK masses of $A_{\mu}$}

For the four-dimensional gauge field $A_{\mu}$, its kinetic term and KK mass term are obtained from the terms
\be
\label{FF}
L=\int d \Omega Tr \biggl[ -\frac{1}{4}F_{\mu \nu} + \frac{1}{2 R^2} F_{\mu \theta} F^{\mu}_{\ \theta}+\frac{1}{2 R^2 \sin^2 \theta} F_{\mu \phi} F^{\mu}_{\ \phi} \biggr] ~.
\ee
Taking terms quadratic in $A_{\mu}$, we get 
\bea
L_{\rm quad} 
&=& \int d \Omega Tr \biggl[ -\frac{1}{4}(\partial_{\mu} A_{\nu}-\partial_{\nu}A_{\mu} )(\partial^{\mu}A^{\nu}-\partial^{\nu}A^{\mu} ) 
+\frac{1}{2 R^2} \partial_{\theta} A_{\mu} \partial_{\theta} A^{\mu} 
\nonumber \\
&& \qquad \qquad 
+ \frac{1}{2 R^2 \sin^2 \theta} \partial_{\phi} A_{\mu} \partial_{\phi} A^{\mu} -\frac{1}{2 R^2} [A_{\mu}, \tilde{A}_{\phi}^B][A^{\mu},\tilde{A}_{\phi}^B] \biggr] ~,
\eea
where $\tilde{A}^B_{\phi}$ is the background gauge field given in Eq.~(\ref{background}).  The KK expansion of $A_{\mu}$ is
\be
A_{\mu} = \sum_{\ell m} A_{\mu}^{\ell m}(x) Y_{\ell m}^{\pm}(\theta,\phi)
\ee
where $Y_{\ell m}^{\pm}(\theta,\phi)$ are the linear combinations of spherical harmonics satisfying the boundary condition $Y_{\ell m}^{\pm}(\pi-\theta,-\phi) = \pm Y_{\ell m}^{\pm}(\theta,\phi)$.  Their explicit forms are \cite{Maru:2009wu}
\bea
\label{modef1}
Y_{\ell m}^+(\theta, \phi) 
&\equiv&  \frac{(i)^{\ell+m}}{\sqrt{2}}[Y_{\ell m}(\theta, \phi) 
+ (-1)^{\ell} Y_{\ell-m}(\theta, \phi)] 
\quad \textrm{for} \quad m \not=0  \\
\label{modef2}
Y_{\ell m}^-(\theta, \phi) 
&\equiv&  \frac{(i)^{\ell+m+1}}{\sqrt{2}}[Y_{\ell m}(\theta, \phi) 
- (-1)^{\ell} Y_{\ell-m}(\theta, \phi)] 
\quad \textrm{for} \quad m \not=0  \\
\label{modef3}
Y_{\ell0}^{+(-)}(\theta) 
&\equiv&
\left\{\begin{array}{l}
Y_{\ell0}(\theta) \quad \textrm{for} \quad m=0 \ \textrm{and} \ 
\ell=\textrm{even (odd)} \\
0 \qquad \quad \textrm{for} \quad m=0 \ \textrm{and} \ 
\ell=\textrm{odd (even)}.
\end{array}\right. 
\eea
Note that we do not have KK mode functions that are odd under $\phi \rightarrow \phi + 2 \pi$ since the KK modes are specified by the integer angular momentum quantum numbers $\ell$ and $m$ of gauge field $A_M$ on the two-sphere.  Thus, the components of $A_{\mu}$ and $A_{\theta,\phi}$ with $(+,-)$ or $(-,+)$ parities do not have corresponding KK modes.
Applying the KK expansion and integrating about $d \Omega$, we obtain the kinetic and KK mass terms for the KK modes of $A_{\mu}$
\bea
\label{masstermAm}
L_M 
&=& -\frac{1}{2} \left[
\partial_{\mu} A^{\ell m}_{\nu}(x)-\partial_{\nu}A^{\ell m}_{\mu}(x)
\right]
\left[
\partial^{\mu}A^{\ell m \nu}(x)-\partial^{\nu}A^{\ell m \mu}(x)
\right] 
 + \frac{\ell(\ell+1)}{R^2} A_{\mu}^{\ell m}(x)  A^{\ell m \mu}(x)  \nonumber \\
&& \qquad + \frac{9 q_Z^2}{R^2} \biggl[ \int d \Omega \frac{(\cos \theta \pm 1)^2}{\sin^2 \theta} (Y_{\ell m}^{\mp})^2 \biggr] A_{\mu}^{\ell m}(x) A^{\ell m \mu}(x) ~,
\eea
where we have used $Tr[Q_i Q^i]=2$ and $[A_{\mu}(x),Q_Z] = q_Z (A_{\mu}^i(x) Q_i - A_{i \mu}(x) Q^i )$.  Therefore, the KK masses of $A_\mu$ are 
\bea
\label{KKmass-gauge}
M_{\ell m}^{KK}(A_\mu) &=& 
\frac{1}{R} \sqrt{\ell(\ell+1)+(m^B_{\ell m})^2} ~, \\
(m^B_{\ell m})^2 &=& 
9 q_Z^2 \int d \Omega 
\frac{(\cos \theta \pm 1)^2}{\sin^2 \theta} (Y_{\ell m}^{\mp})^2 ~,
\eea
where $m^B_{\ell m}$ corresponds to the contribution from the background gauge field.  Note that Eq.~(\ref{KKmass-gauge}) agrees with Eq.~(\ref{eq:nonSMgaugeMass}) when $\ell = 0$.  Also, since the SM gauge bosons have $q_Z = 0$, their KK masses are simply $\sqrt{\ell(\ell+1)}/R$ at the tree level.

\subsection{KK masses of $A_{\theta,\phi}$}

The kinetic and KK mass terms of $A_{\theta}$ and $A_{\phi}$ are obtained from the terms in the higher dimensional gauge sector
\bea
\label{scalar}
L &=& \frac{1}{2 g^2} \int d \Omega \Biggl\{ \Bigl( Tr[(\partial_{\mu} A_{\theta}-i[A_{\mu},A_{\theta}])^2] 
+ Tr[(\partial_{\mu} A_{\theta}-i[A_{\mu},\tilde{A}_{\phi}])^2 ]  \Bigr) \nonumber \\
&& \qquad \qquad -  \frac{1}{R^2} Tr \biggl[  \biggl( \frac{1}{\sin \theta} \partial_{\theta} (\sin \theta \tilde{A}_{\phi} 
+ \sin \theta \tilde{A}^B_{\phi}) -\frac{1}{\sin \theta} \partial_{\phi} A_{\theta} 
- i[A_{\theta},\tilde{A}_{\phi}+\tilde{A}^B_{\phi}] \biggr)^2 \biggr] \Biggr\} ~. \nonumber \\
\eea
The first line on the right-hand side of Eq.~(\ref{scalar}) corresponds to the kinetic terms, and the second line corresponds to the potential term.  Applying the background gauge field Eq.~(\ref{background}), the potential becomes
\bea
L_V = -\frac{1}{2 g^2 R^2} \int d \Omega Tr \biggl[ 
\biggl( \frac{1}{\sin \theta} \partial_{\theta} (\sin \theta \tilde{A}_{\phi}) + Q - \frac{1}{\sin \theta} \partial_{\phi} A_{\theta} -i [A_{\theta}, \tilde{A}_{\phi}+\tilde{A}_{\phi}^B] \biggr)^2 
\biggr]
\eea
For $A_{\theta}$ and $A_{\phi}$ we use the following KK expansions to obtain the KK mass terms,
\bea
\label{expansion3}
A_{\theta}(x,\theta,\phi) 
&=& \sum_{\ell m (\neq 0)} \frac{-1}{\sqrt{\ell(\ell+1)}} \bigl[ \Phi_1^{\ell m}(x) \partial_{\theta} Y_{\ell m}^{\pm}(\theta,\phi) 
+ \Phi_2^{\ell m}(x) \frac{1}{\sin \theta} \partial_{\phi} Y_{\ell m}^{\pm}(\theta,\phi)  \bigr] ~, \\
\label{expansion4}
A_{\phi}(x,\theta,\phi) 
&=& \sum_{\ell m (\neq 0)}\frac{1}{\sqrt{\ell(\ell+1)}} 
\bigl[ \Phi_2^{\ell m}(x) \partial_{\theta} Y_{\ell m}^{\pm}(\theta,\phi)
  - \Phi_1^{\ell m}(x) \frac{1}{\sin \theta} \partial_{\phi} Y_{\ell m}^{\pm}(\theta,\phi)  \bigr] ~,
\eea
where the factor of $1/\sqrt{\ell(\ell+1)}$ is needed for normalization.  These particular forms are convenient in giving diagonalized KK mass terms \cite{Maru:2009wu}.  Applying the KK expansions Eq.~(\ref{expansion3}) and Eq.~(\ref{expansion4}), we obtain the kinetic term
\bea
L_{K} = \frac{1}{2 g^2} \sum_{\ell m (\neq 0)} Tr \biggl[ \partial_{\mu} \Phi_1^{\ell m}(x) \partial^{\mu} \Phi_1^{\ell m}(x) +  \partial_{\mu} \Phi_2^{\ell m}(x) \partial^{\mu} \Phi_2^{\ell m}(x) \biggr]
\eea
where only terms quadratic in $\partial_{\mu} \Phi$ are retained.  The potential term
\bea
L_V 
&=& -\frac{1}{2 g^2 R^2} \sum_{\ell m (\neq 0)} \int d \Omega 
Tr \biggl[  \biggl( \frac{\Phi_{2}^{\ell m}}{\sqrt{\ell(\ell+1)}} 
\frac{1}{\sin \theta} \partial_{\theta} 
(\sin \theta \partial_{\theta} Y_{\ell m}^{\pm} ) +Q 
+ \frac{\Phi_{2}^{\ell m}}{\sqrt{\ell(\ell+1)}} \frac{1}{\sin^2 \theta} \partial_{\phi}^2 Y_{\ell m}^{\pm} \nonumber \\ 
&& \qquad
- \frac{i}{\ell(\ell+1)} \Bigl[ - \Phi_{1}^{\ell m} \partial_{\theta} Y_{\ell m}^{\pm} - \Phi_{2}^{\ell m} \frac{1}{\sin \theta} \partial_{\phi} Y_{\ell m}^{\pm}, 
\Phi_{2}^{\ell m} \partial_{\theta} Y_{\ell m}^{\pm} - \Phi_{1}^{\ell m} \frac{1}{\sin \theta} \partial_{\phi} Y_{\ell m}^{\pm} 
\nonumber \\
&& \qquad
+ \sqrt{\ell(\ell+1)} A_{\phi}^B \Bigr] \biggr)^2 \biggr] ~,
\nonumber \\
\eea
where only terms diagonal in $(\ell,m)$ are consider.  Using the relation $\frac{1}{\sin \theta} \partial_{\theta} (\sin \theta \partial_{\theta} Y_{\ell m}) + \frac{1}{\sin^2 \theta} \partial_{\phi}^2 Y_{\ell m} = -\ell(\ell+1)Y_{\ell m}$, the potential term is simplified as
\bea
L_V &=& -\frac{1}{2 g^2 R^2} \sum_{\ell m(\neq 0)} \int d \Omega Tr \biggl[ \biggl( 
-\sqrt{\ell(\ell+1)}  \Phi_2^{\ell m}Y_{\ell m}^{\pm} +Q \nonumber \\
&& \qquad \qquad \qquad 
+\frac{i}{\ell(\ell+1)} [\Phi_1^{\ell m}, \Phi_2^{\ell m}]  \bigl( \partial_{\theta} Y_{\ell m}^{\pm} \partial_{\theta} Y_{\ell m}^{\pm}
+\frac{1}{\sin^2 \theta} \partial_{\phi} Y_{\ell m}^{\pm} \partial_{\phi} Y_{\ell m}^{\pm} \bigr) \nonumber \\
&& \qquad \qquad \qquad 
+\frac{i}{\sqrt{\ell(\ell+1)}} [\Phi_1^{\ell m}, \tilde{A}_{\phi}^B] \partial_{\theta} Y_{\ell m}^{\pm} 
+ \frac{i}{\sqrt{\ell(\ell+1)}} [\Phi_2^{\ell m}, \tilde{A}_{\phi}^B] \frac{\partial_{\phi} Y_{\ell m}^{\pm}}{\sin \theta} \biggr)^2 \biggr] ~.
\eea
To obtain the mass term, we focus on terms quadratic in $\Phi_{1,2}$: 
\bea
\label{massterm}
L_M &=& -\frac{1}{2 g^2 R^2} \int d \Omega Tr \biggl[ 
\ell(\ell+1) (\Phi_2^{\ell m})^2 (Y_{\ell m}^{\pm})^2 \nonumber \\
&& \qquad
+ \frac{2i Q}{\ell(\ell+1)} [\Phi_1^{\ell m},\Phi_2^{\ell m}] 
\Bigl( \partial_{\theta} Y_{\ell m}^{\pm} \partial_{\theta} Y_{\ell m}^{\pm} + \frac{1}{\sin^2 \theta} \partial_{\phi} Y_{\ell m}^{\pm} \partial_{\phi} Y_{\ell m}^{\pm} \Bigr) \nonumber \\
&& \qquad
+2 i \tilde{A}_{\phi}^B [\Phi_1^{\ell m},\Phi_2^{\ell m}] Y_{\ell m}^{\pm} \partial_{\theta} Y_{\ell m}^{\pm} 
- \frac{1}{\ell(\ell+1)} [\Phi_1^{\ell m},\tilde{A}_{\phi}^B]^2 (\partial_{\theta} Y_{\ell m}^{\pm})^2 \nonumber \\
&& \qquad
- \frac{1}{\ell(\ell+1)} [\Phi_2^{\ell m},\tilde{A}_{\phi}^B]^2 \frac{(\partial_{\phi} Y_{\ell m}^{\pm})^2}{\sin^2 \theta} \biggr]. \nonumber \\
\eea
Note that we have dropped the term proportional to $[\Phi_1,\tilde{A}_{\phi}^B] [\Phi_2,\tilde{A}_{\phi}^B]$ because this term vanishes after turning the field into the linear combinations of $\Phi$ and $\Phi^\dagger$, Eqs.~(\ref{okikae1}) and (\ref{okikae1}): 
\bea
Tr[[\Phi_1,\tilde{A}_{\phi}^B][\Phi_1,\tilde{A}_{\phi}^B]] &\rightarrow& Tr[[(\Phi+\Phi^{\dagger}),Q] [(\Phi-\Phi^{\dagger}),Q] ] \nonumber \\
&\propto& Tr[(\Phi-\Phi^\dagger)(\Phi+\Phi^{\dagger})] \nonumber \\
&\propto& Tr[\Phi \Phi^{\dagger}] - Tr[\Phi^{\dagger} \Phi] =0
\eea
Integrating the second term of Eq.~(\ref{massterm}) by part, we obtain
\bea
L_M &=& -\frac{1}{2g^2 R^2} \biggl( \ell(\ell+1) Tr[(\Phi_2^{\ell m})^2] +2i Tr[Q [\Phi_1^{\ell m},\Phi_2^{\ell m}] ] \nonumber \\
&& \qquad \qquad  -2 i Tr[Q [\Phi_1^{\ell m},\Phi_2^{\ell m}] ]  \int d \Omega \frac{\cos \theta \mp 1}{\sin \theta} Y_{\ell m}^{\pm} \partial_{\theta} Y_{\ell m}^{\pm} \nonumber \\
&& \qquad \qquad -\frac{1}{\ell(\ell+1)} [\Phi_1^{\ell m},Q]^2 \int d \Omega \frac{(\cos \theta \mp1)^2}{\sin^2 \theta} (\partial_{\theta} Y_{\ell m}^{\pm})^2 \nonumber \\
&& \qquad \qquad - \frac{1}{\ell(\ell+1)} [\Phi_2^{\ell m},Q]^2 \int d \Omega \frac{(\cos \theta \mp 1)}{\sin^2 \theta} \frac{(\partial_{\phi} Y_{\ell m}^{\pm})^2}{\sin^2 \theta} \biggr) ~. \nonumber \\
\eea
Therefore, the KK masses depend on the U(1)$_Z$ charges of the scalar fields.

For components with zero U(1)$_Z$ charge, we write $\Phi_{1(2)}(x)$ as $\phi_{1(2)}(x) Q$ where $Q$ is the corresponding generator of E$_6$ in Eq.~(\ref{d78}) with zero U(1)$_Z$ charge.  Taking the trace, we have the following kinetic and KK mass terms instead:
\bea
L =  \sum_{\ell m(\neq 0)} \biggl( \partial_{\mu} \phi_1^{\ell m}(x) \partial^{\mu} \phi_1^{\ell m}(x) +  \partial_{\mu} \phi_2^{\ell m}(x) \partial^{\mu} \phi_2^{\ell m}(x)  
+ \ell(\ell+1) \phi_2^{\ell m}(x) \phi_2^{\ell m}(x) \biggr)
\eea
where we have made the substitution $\phi_i \rightarrow g \phi_i$.  Note that $\phi_1$ is considered as a massless Nambu-Goldstone (NG) boson in this case.

For components with nonzero U(1)$_Z$ charge, we use Eq.~(\ref{okikae1}) and (\ref{okikae2}) and write $\Phi(x)$ as $\phi^i(x)Q_i$ where $Q_i$ is the corresponding generator of $E_6$ in Eq.~(\ref{d78}) with nonzero U(1)$_Z$ charge.  The commutator between $Q$ and $\Phi$ is
\be
[Q,\Phi] = 3[Q_Z,Q_i] \phi^i = 3 q_Z \phi^i ~,
\ee
where we have used $Q=3Q_Z$ and that $q_Z$ is a constant determined by the U(1)$_Z$ charge of the corresponding component.  Finally, the Lagrangian becomes
\bea
L &=&  \sum_{\ell m(\neq 0)} \biggl\{ \partial_{\mu} \phi^{\dagger}_{\ell m}  \partial^{\mu} \phi_{\ell m} \nonumber \\
&& \qquad \quad -\frac{1}{4 R^2} \biggl[ 
2 \ell(\ell+1) \phi_{\ell m}^\dagger \phi_{\ell m}
-12 q_Z \phi_{\ell m}^\dagger \phi_{\ell m} 
+12 q_Z \phi_{\ell m}^\dagger \phi_{\ell m } 
\int d \Omega \frac{\cos \theta \mp 1}{\sin \theta} Y_{\ell m}^{\pm} \partial_{\theta} Y_{\ell m}^{\pm} \nonumber \\
&& \qquad \qquad \qquad 
+\frac{18 q_Z^2}{ \ell(\ell+1)} \phi_{\ell m}^\dagger \phi_{\ell m} 
\int d \Omega \frac{(\cos \theta \mp1)^2}{\sin^2 \theta} 
\left(
(\partial_{\theta} Y_{\ell m}^{\pm})^2
+ \frac{(\partial_{\phi} Y_{\ell m}^{\pm})^2}{\sin^2 \theta}
\right) \biggr] \biggr\}. \nonumber \\
\eea
where the subscript $i$ is omitted for simplicity.  The KK masses of the complex scalar field $\phi$ are then
\begin{eqnarray}
\label{KKmass-scalar}
M_{\ell m}^{KK}(\phi) &=& \frac{1}{R} \sqrt{\frac{\ell(\ell+1)}{2}+(m_{\ell m}^B)^2} ~, \nonumber \\
(m_{\ell m}^B)^2 &=& -3 q_Z  +3 q_Z  \int d \Omega \frac{\cos \theta \mp 1}{\sin \theta} Y_{\ell m}^{\pm} \partial_{\theta} Y_{\ell m}^{\pm} 
+\frac{9 q_Z^2}{ 2\ell(\ell+1)} \int d \Omega \frac{(\cos \theta \mp1)^2}{\sin^2 \theta} (\partial_{\theta} Y_{\ell m}^{\pm})^2 \nonumber \\
&& \qquad + \frac{9 q_Z^2}{2\ell(\ell+1)}  \int d \Omega \frac{(\cos \theta \mp 1)^2}{\sin^2 \theta} \frac{(\partial_{\phi} Y_{\ell m}^{\pm})^2}{\sin^2 \theta} ~.
\end{eqnarray}
The squared KK mass $\left( M_{\ell m}^{KK} \right)^2$ is always positive except for the lowest mode ($\ell=1,m=1$).  In fact, the squared KK mass of the $(1,1)$ mode agrees with the coefficient of quadratic term in the Higgs potential (\ref{H-potential}).

\subsection{Dark matter candidate}

In our model, each KK particle is associated with a KK parity derived from an additional $Z_2'$ discrete symmetry of $(\theta,\phi) \rightarrow (\theta,\phi + \pi)$, corresponding to the exchange of the two fixed points on the orbifold \cite{Maru:2009wu}.  The KK-parity is given by $(-1)^m$, and is conserved as a consequence of the $Z_2'$ symmetry of the Lagrangian in six-dimensional spacetime.  Therefore, the lightest KK particle with an odd $m$ will be stable.

A comparison among Eqs.~(\ref{KKmass-fermion}), (\ref{KKmass-gauge}) and (\ref{KKmass-scalar}) indicates that the lightest KK particles are the $(\ell=1,m=1)$ modes of the scalar components with non-zero U(1)$_Z$ charges since their masses receive a negative contribution from the background gauge field.  They include the components $\{(3,2)(1,-1,-3)^{(-,-)} + {\rm h.c.} \}$ and $\{(1,2)(-3,3,-3)^{(+,+)} + {\rm h.c.} \}$ in Eq.~(\ref{78scalar}) since the other components either have zero U(1)$_Z$ charge or are odd under $\phi \rightarrow \phi + 2 \pi$.  At the tree level, both of them have the same and negative KK squared mass since their U(1)$_Z$ charges are same ($q_Z = 1/2$).  As argued at the end of Section~\ref{sec:Higgssector}, the squared mass of the former representation can be lifted by brane terms to be sufficiently large to avoid color symmetry breaking.  Its mass depends on the parameter $\alpha$ in the brane terms.  The latter representation actually gives the Higgs field that has a mass about 152 GeV.  We assume that the mass of the $(1,1)$ mode of the $\{(3,2)(1,-1,-3)^{(-,-)} + {\rm h.c.} \}$ components is heavier than the Higgs boson mass since a colored particle is not suitable for dark matter candidate.  Therefore, the model has the $(1,1)$ mode of the $\{(1,2)(-3,3,-3)^{(+,+)} + {\rm h.c.} \}$ representation as the lightest and stable KK particle, which is simply the Higgs boson.

\section{Summary \label{sec:summary}}

The gauge-Higgs unification is an attractive idea because it can unify the SM gauge bosons and Higgs boson under a higher dimensional spacetime symmetry.  The gauge invariance prevents the Higgs boson in the bulk from receiving radiative corrections that diverge with the cutoff scale, thus easing the gauge hierarchy problem.  However, one still encounters the difficulty in getting an appropriate Higgs potential to break the electroweak symmetry in five dimensional spacetime.  Extra particles are generally needed in order to generate such a potential and a sufficiently large Higgs mass radiatively.  When one goes to six spacetime dimensions and considers the $S^2/Z_2$ orbifold, it is possible to render a suitable Higgs potential by incorporating a background gauge field in the extra-dimensional components.  To fully achieve that, nevertheless, one has to assume a special symmetry that relates the SU(2) isometry transformation in $S^2$ to the gauge transformation.

We consider in this paper a six-dimensional gauge-Higgs unification model, in which the gauge group is enlarged to E$_6$ and the extra space is the $S^2/Z_2$ orbifold.  By specifying two sets of parity transformation properties for the fields and employing a Dirac monopole configuration for the background gauge field, we have a successful symmetry reduction to the SM gauge group plus two extra U(1)'s.  In our model, the background gauge field $A_{\phi}^B$ plays important roles in several aspects.  First, it renders massless chiral fermions by canceling the spin-connection term in the covariant derivative.  Secondly, it elevates the masses of unwanted representations of $A_\mu$ to roughly the compactification scale in four dimensional spacetime.  Finally, from the gauge kinetic term, it gives rise to a negative mass square term for the Higgs potential at tree level.

At the low energy, we obtain only the SM particles.  The SM gauge bosons all originate from a single adjoint representation of the E$_6$ group.  The chiral fermions, including a right-handed neutrino, are derived from four copies of fundamental representation, each of which have a distinct parities under the two parity transformations.  We also obtain exactly one complex Higgs doublet from the extra dimensional components of the gauge field.

We have computed the Higgs potential in this model.  The squared mass is related to the compactification radius, and the quartic coupling to the E$_6$ gauge coupling.  The radius of the compactified two-sphere is extracted to be around (152 GeV)$^{-1}$.  The Higgs boson mass is predicted to be about 150 GeV at tree level.  Due to the gauge group structure, we obtain $\sin^2\theta_W = 3/8$, the same as in the SU(5) GUT at the unification scale.

Through KK expansions, we have calculated the mass spectra of the gauge and fermion fields.  In general, these masses involve two contributions: one related to the angular momentum eigenvalues in the extra dimensions $\ell(\ell+1)$, and the other due to the interactions between the KK modes and the background gauge field.  Finally, the model can have a dark matter candidate due to the KK parity under the $Z'_2$ symmetry.  It is the Higgs boson of the model.

\section*{Acknowledgments}

This research was supported in part by the National Science Council of Taiwan,
R.O.C.\ under Grant No.~NSC~97-2112-M-008-002-MY3 and the NCTS.

\end{document}